\documentclass[%
 reprint,
 amsmath,amssymb,
 aps,
longbibliography,
]{revtex4-1}

\usepackage{graphicx}% Include figure files
 \usepackage{float}

\usepackage{dcolumn}% Align table columns on decimal point
\usepackage{bm}% bold math
\begin{document}
\newcommand{\sech}{\mathrm{sech}}

\preprint{APS/123-QED}

\title{Spontaneous and engineered  transformations of  topological structures in nonlinear media with gain and loss}
\author{B.~A.~Kochetov$^1$}
\author{O.~G.~Chelpanova$^{2}$}
\author{V.~R.~Tuz$^{1,3}$}
\email{tvr@jlu.edu.cn; tvr@rian.kharkov.ua}
\author{A.~I.~Yakimenko$^{2}$}
\affiliation{$^1$State Key Laboratory of Integrated Optoelectronics, College of Electronic Science and Engineering,\\ International Center of Future Science, Jilin University, 2699 Qianjin Street, Changchun 130012, China}
\affiliation{$^2$Department of Physics, Taras Shevchenko National University of Kyiv, 64/13, Volodymyrska Street, Kyiv 01601, Ukraine}
\affiliation{$^3$Institute of Radio Astronomy, National Academy of Sciences of Ukraine, 4, Mystetstv Street, Kharkiv 61002, Ukraine}
%\date{\today}

\begin{abstract}
In contrast to conservative systems, in nonlinear media with gain and loss the dynamics of localized topological structures can exhibit unique features that can be controlled externally. We propose a robust mechanism to perform topological transformations changing characteristics of dissipative vortices and their complexes in a controllable way. We show that a properly chosen control carries out the evolution of dissipative structures to regime with spontaneous transformation of the topological excitations or drives generation of vortices with control over the topological charge.
\end{abstract}

%\pacs{05.45.-a, 05.45.Yv, 42.65.Tg, 42.65.Wi}
%05.45.-a	Nonlinear dynamics and chaos
%05.45.Yv	Solitons
%42.65.Tg	Optical solitons; nonlinear guided waves
%42.65.Wi	Nonlinear waveguides

\maketitle

% \tableofcontents
%%%%%%%%%%%%%%%%%%%%%%%%%%%%%%%%%%%%%%%%
%\section{\label{sec:intro}Introduction}
%%%%%%%%%%%%%%%%%%%%%%%%%%%%%%%%%%%%%%%

Formation of vortices, excitations that possess a rotational flow around a point of phase singularity, is a general phenomenon observed in various fields of both classical and quantum physics, including acoustics, fluid dynamics, solid-state physics, Bose-Einstein condensation, etc. \cite{Pismen_Book}. In optics the unprecedented attention has been attracted to the vortex formation in nonlinear optical systems, where linear spreading due to the dispersion or diffraction is balanced by nonlinear focusing \cite{Desyatnikov_PO_2005, Soskin_JO_2017}. Nowadays vortices are a subject of numerous studies in nonequilibrium polariton condensates \cite{Lagoudakis_NPh_2008, Lagoudakis_S_2009, Ostrovskaya_PRA_2012, Dall_PRL_2014, Ma_PRL_2017, Sigurdsson_PRB_2014, Ma_PRB_2016,  Ma_PRL_2018}. Besides the studies on vortex existence and stability \cite{Ostrovskaya_PRA_2012, Ma_PRL_2017} as well as vortex-antivortex dynamics \cite{Fraser_NJP_2009, Boulier_CRP_2016}, tremendous efforts have been aimed to get the full control over the formation of vortices with required topological properties using a broad optical pump \cite{Lagoudakis_NPh_2008, Lagoudakis_S_2009}, chiral polaritonic lenses \cite{Dall_PRL_2014}, and a ring-shaped incoherent optical pump \cite{Sigurdsson_PRB_2014, Ma_PRB_2016, Ma_PRB_2017, Ma_PRL_2018}. Formation in a predefined way of stable vortices with an arbitrary topological charge from a small-amplitude noise is possible, when an incoherent pump is locally applied to the exciton-polariton condensate \cite{Ma_PRB_2016}. Moreover, the selective transfer of topological charge between two vortices with unit positive and negative charges has been realized in the condensate due to the presence of an elliptically-shaped incoherent control beam \cite{Ma_PRB_2016}. Further elaboration of this mechanism based on controlling vortex multistability has allowed to switch a topological state with the charge $m=\pm1$, $\pm2$, $\pm3$ to another vortex \cite{Ma_PRL_2018}.

Despite significant progress in the vortex formation in diverse physical systems, the getting control over vorticity of multi-dimensional dissipative solitons in a predefined way is still a challenging problem for many hydrodynamic and optical applications. In this work we propose a mechanism to perform nontrivial topological transformations by applying an external potential which changes topological characteristics of dissipative vortices in a controllable way. We demonstrate this mechanism in the framework of the complex Ginzburg-Landau equation (CGLE) with a potential term, which covers nonlinear phenomena far from equilibrium in many physical systems including mode-locked and fiber lasers, nonlinear optical waveguides, semiconductor devices, Bose-Einstein condensates, reaction-diffusion systems, etc. \cite{Cross_RMP_1993, Aranson_RMP_2002, Akhmediev_Book1, Akhmediev_Book2, Garcia-Morales_CP_2012, Liehr_Book}. We show that the properly chosen potential performs preassigned transformations of topological structures.

In the chosen framework, having adopted the notations used in the theory of nonlinear optical waveguides \cite{Boardman_Chapter_2005, Boardman_2006, OptLett_2017} and active optical media \cite{He_OL_2009, Leblond_PRA_2009, Liu_OL_2010, He_JOSAB_2010, Yin_JOSAB_2011, Liu_OE_2013}, we write the cubic-quintic CGLE supplemented by a potential term with an explicit coordinate dependence as follows
\begin{multline}
\label{eq:CQCGLE}
\mathrm{i}\frac{\partial\Psi}{\partial z}+\mathrm{i}\delta\Psi+\left(\frac{D}{2}-\mathrm{i}\beta\right)\nabla^2\Psi
+\left(1-\mathrm{i}\varepsilon\right)\left|\Psi\right|^2\Psi\\ -\left(\nu-\mathrm{i}\mu\right)\left|\Psi\right|^4\Psi + Q(x,y,z)\Psi=0,
\end{multline}
where $\Psi\left(x,y,z\right)$ is the slowly varying vortex envelop, which is the complex function of two transverse $(x$ and $y)$ and the longitudinal $(z)$ coordinates. The coefficients of Eq.~\eqref{eq:CQCGLE} are assumed to be positive quantities that results in their unambiguous interpretation. Namely, $D$ is the normalized diffraction coefficient, $\delta$ is the linear absorption coefficient, $\beta$ is the linear diffusion coefficient, $\varepsilon$ is the coefficient of nonlinear cubic gain, $\nu$ accounts for the self-defocusing effect, and $\mu$ defines quintic nonlinear losses. The potential $Q(x,y,z)$ is a real function, which accounts for a specific conservative force applied externally that influences over the evolution of complex envelop $\Psi(x,y,z)$. It is given in the form
\begin{equation}
\label{eq:Q}
Q(x,y,z)=\sum_{i=1}^{N_Q}q_{i}(x,y,z)\left[\theta(z-a_{i})-\theta(z-b_{i})\right],
\end{equation}
where $N_Q$ is the number of control manipulations, $\theta$ is the Heaviside step function, and $a_{i}<b_{i}$ are the points on the $z$ axis where the potential is switched between different states.

For a given set of coefficients taken from relatively wide ranges of values, Eq.~\eqref{eq:CQCGLE} with zero potential has numerous stable solutions in the form of dissipative vortices with different topological charges. Each vortex corresponds to certain attractor in the phase space of Eq.~\eqref{eq:CQCGLE} with zero potential. To excite the vortex one can use an arbitrary initial condition, which starts a phase trajectory somewhere within the basin of attraction. In all our simulations we use the same fixed set of coefficients of Eq.~\eqref{eq:CQCGLE}: $D=1$, $\beta=0.5$, $\delta=0.5$, $\mu=1$, $\nu=0.1$, and $\varepsilon=2.5$. For this set of coefficients one can use the following initial condition to excite a given vortex
\begin{equation}
\label{eq:IC}
\Psi_0(x,y)=A_m \exp\left(i m\varphi -\frac{r^2}{w^2}\right),
\end{equation}
where $\varphi$ is the azimuthal angle, the amplitude $A_m=1$ and effective radius $w=3$. The integer $m$ defines the vortex topological charge, which corresponds to the circulation of the gradient of the phase on a closed curve surrounded the vortex core. In our simulations we use $m=0$, $m=1$, and $m=-1$ which correspond to three different cases when we initially excite the fundamental soliton, vortex, and antivortex with unit topological charges, respectively.

We solve Eq.~\eqref{eq:CQCGLE} numerically using the exponential time differencing method and its two Runge-Kutta modifications of the second and fourth order accuracy \cite{Cox_2002} as well as the split-step Fourier method of the second order accuracy \cite{Agrawal_NWO}.

Being formed a vortex cannot change its waveform as long as parameters in Eq.~\eqref{eq:CQCGLE} are fixed. On the other hand, having applied the nonzero potential we can perturb the vortex pushing out its phase trajectory from the original basin of attraction to another one. As soon as the basin of attraction has been changed we switch off the potential and the released waveform evolves to a new appearance. Such mechanism of induced waveform transitions between different one-dimensional dissipative solitons has been thoroughly studied in Refs.~\onlinecite{Chaos_2018, PD_2019}. Here our goal is to show that this mechanism is quite general and provides tremendous opportunities to control the vortex formation and perform transformations between their different topological charges. Further we distinguish two different mechanisms related to spontaneous and engineered transformations.

It was previously revealed \cite{Chaos_2018} that a one-dimensional nonlinear system with gain and loss can exhibit chaotic behavior being driven by an external potential. Nevertheless, somewhat control over the finite state is possible by choosing the position of switching off the external potential to release the soliton from chaotic behavior to stationary one. This mechanism is related to spontaneous transformations of solitons.

In order to illustrate such transformations for vortices we consider a single manipulation potential ($N_Q=1$). This potential is homogeneous in the $y$-direction and constructed as a superposition of two potential wells with minima at $x=\pm x_0$ separated by potential barrier with $x=0$ and $x_0=10$:
\begin{equation}
\label{eq:spontaneous}
 q_{1}(x)=\sech(x-x_0)-\sech(x)+\sech(x+x_0).
\end{equation}

To monitor the appearance of the topological excitations we use the core detection technique \cite{caradoc2000vortex}, which allows us to determine a proper value of $b_1$ when the potential should be switched off. For the same purpose we also calculate energy $E=\int |\Psi|^2 d^2 \mathbf{r}$, angular momentum $\mathbf{L}=-\frac{i}{2}\int \left[\mathbf{r}\times(\Psi^{*} \nabla\Psi-\Psi\nabla\Psi^{*})\right]d^2\mathbf{r}$, and momentum of inertia $I=\int |\Psi|^2r^2d^2\mathbf{r}$, where $r=\sqrt{x^2+y^2}$ is the radial coordinate. In general, energy $E$, angular momentum $\mathbf{L}$, and momentum of inertia $I$ are functions of the $z$ coordinate rather than conserved quantities. However, for any localized structure these characteristics are always finite. Moreover, in a stationary case they take constant values suitable for monitoring the dynamics of system \eqref{eq:CQCGLE}.

%In this study we use the core detection technique \cite{caradoc2000vortex} for monitoring the appearance of the topological excitations, which allows us to determine a proper value of $b_1$ when the potential should be switched off.

%We introduce the following characteristics of the system \eqref{eq:CQCGLE}: energy $N=\int |\Psi|^2 d^2 \mathbf{r}$, angular momentum $\mathbf{L}=-\frac{i}{2}\int \left[\mathbf{r}\times(\Psi^{*} \nabla\Psi-\Psi\nabla\Psi^{*})\right]d^2\mathbf{r}$ and momentum of inertia $I=\int |\Psi|^2r^2d^2\mathbf{r}$, where $r=\sqrt{x^2+y^2}$ is the radial coordinate. In general, energy $N$, angular momentum $\mathbf{L}$, and momentum of inertia $I$ are functions of the $z$ coordinate rather than conserved quantities. However, for any localized structure these characteristics are always finite. Moreover, in a stationary case those characteristics take constant values that makes them the convenient quantities to monitor the system dynamics.

For each $z$ we calculate the topological charge of the system: $\Delta m=m_{+}-m_{-}$, where $m_{+}$ and $m_{-}$ are the numbers of vortices and antivortices, respectively, and $L=L_z/E$ is the relative angular momentum. Using the information about topological structures available for different $z$, we predict the range of $b_1$, for which the finite state acquires a particular topological charge.

\begin{figure}[t!]%[htb]
\includegraphics[width=8.5cm]{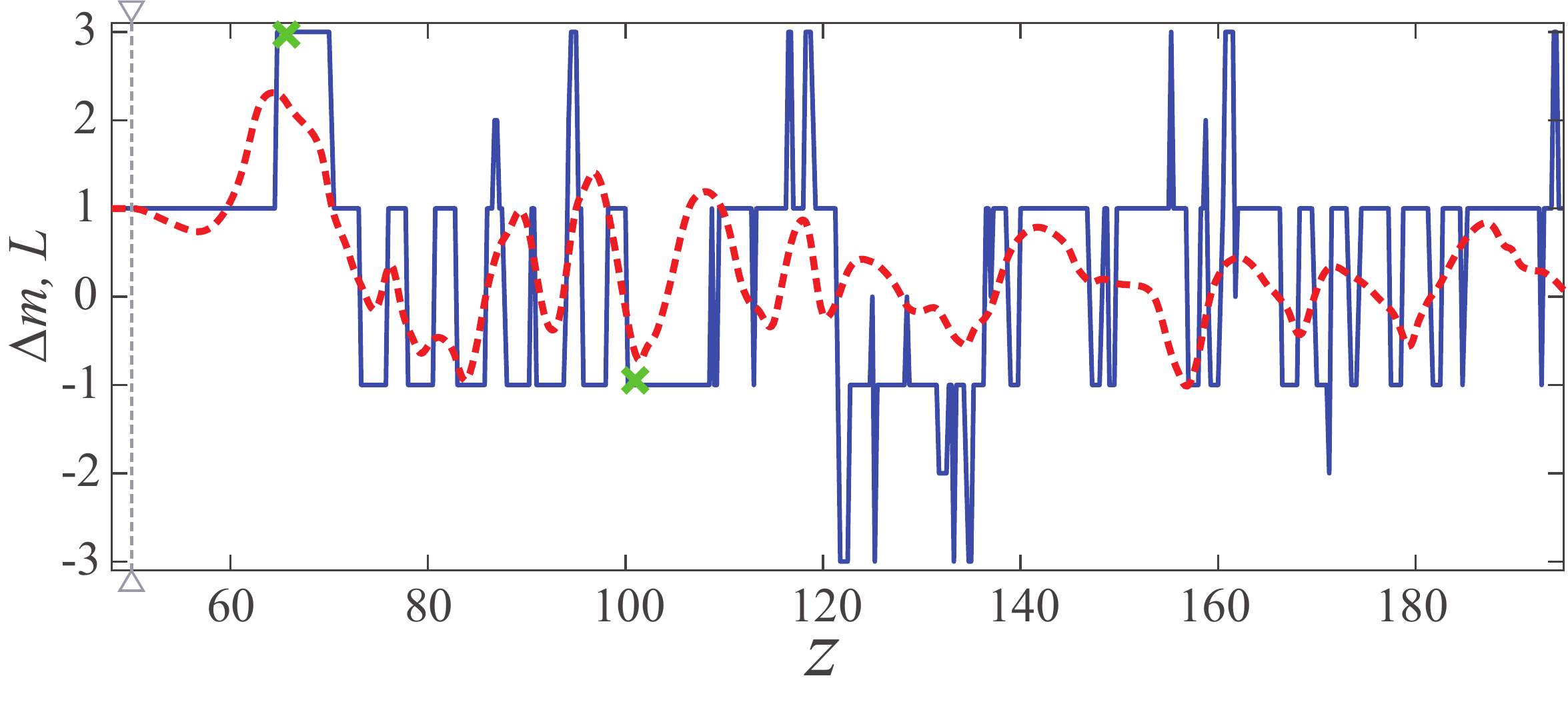}
\caption{\label{fig:L_z} Evolution in $z$-direction of topological charge $\Delta m$ (solid blue line) and angular momentum per particle $L$ (dashed red line). The green crosses mark the position $z=b_1$ where the potential should switched off to form a single-charged dissipative vortex soliton.}
\end{figure}

\begin{figure}[t!]%[htbp]
\centering
\includegraphics[width=8.5cm]{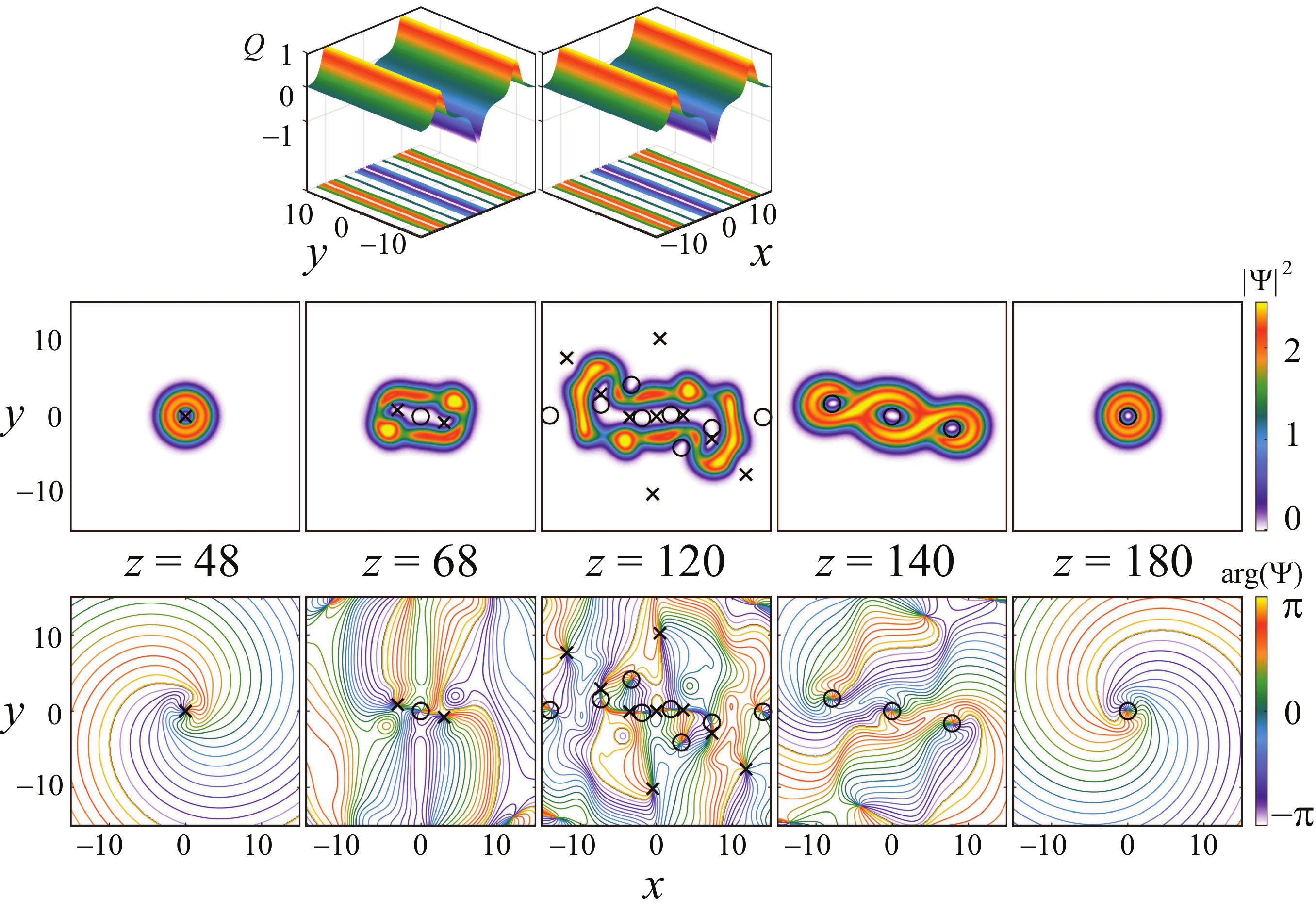}
\caption{Vortex to antivortex transformation obtained with control duration of potential action ($a_1=50$ and $b_1=101$). Snapshots of the spatial distribution of potential (top panel), density (middle panel), and phase (bottom panel) for different $z$. Crosses (circles) mark the vortex (antivortex) cores.} \label{fig:one_to_minus_one}
\end{figure}

As illustrated in Figs.~\ref{fig:L_z} and \ref{fig:one_to_minus_one} in some cases this method allows one to control a type of topological structure which appears after the relaxation process. Figure~\ref{fig:L_z} shows that the number of vortices detected in the wave field changes irregularly, but the analysis of the angular momentum state guides one to choose a proper moment when the potential should be switched off. For instant, it turns out that the manipulation governing by the potential \eqref{eq:spontaneous} with $a_1=50$ and $b_1=101$ transfers charge of vortex from $m=+1$ to $m=-1$. As it is seen from Fig. \ref{fig:L_z} for $z=65.5$ number of detected vortex cores anticipates approaching to the basin  with three-charged vortex structure. Indeed, the same potential \eqref{eq:spontaneous} with $a_1=50$ and $b_1=65.5$ transforms vortex with $m=1$ to the three-vortex rotating cluster, that finally evolves to the three-charged vortex ($m=3$) in good agreement with prediction of the angular momentum analysis (see the Supplemental Material \cite{Suppl_Mat} for animation). A wide variety of different topological excitations can emerge and decay in highly nonequilibrium state. A detailed consideration of dissipative vortex transformation in a regime of strong two-dimensional turbulence may be a relevant extension of this work, which might help to illuminate fundamental properties of the turbulence in classical and quantum physical systems.

Although effective, the mechanism of spontaneous transformations have several drawbacks. It is not always possible to control the final state for irregular evolution induced by simple potential with chosen parameter $b_1$. Also the number of possible transformations is restricted by available topological structures which emerge and decay in the perturbed wave field. In what follows we describe how the potential can be used to obtain a complete control over the mutual transformations of different types of dissipative solitons for formation of complicated topological structures. We relate this mechanism to engineered transformations of dissipative solitons induced by external control.

Due to topological reasons it is possible to get a single vortex only from the periphery of the wave beam, and only vortex-antivortex pairs can be excited inside the wave beam.
The system with gain and loss conserves neither energy nor angular momentum, which allows topological transformation forbidden for conservative systems.  Here we demonstrate how the well known in conservative systems methods of the vortex generation can be modified to generate dissipative vortices. Furthermore, we  suggest novel approaches for creation of topological excitations in the controllable dissipative systems.

A variable in $z$-direction control potential is used to create a  rotating repulsive barrier, similar to stirring wave beam used in Bose-Einstein condensates \cite{wright2013driving, eckel2014hysteresis, yakimenko2015vortices, yakimenko2015vortex}:
\begin{equation}
q_\textrm{st}(x,y,z)=-h  \theta\left(\vartheta(x,y,z)\right)\exp[-\vartheta^2(x,y,z)/r_0^2],
\end{equation}
where $r_0$ is the width of the stirrer, $h$ is the depth of potential, $\vartheta(x,y,z)=y\cos \Omega z-x\sin\Omega z.$  Here $\Omega$ is the angular velocity of the stirrer.

\begin{figure}[t!]
\centering
\includegraphics[width=8.5cm]{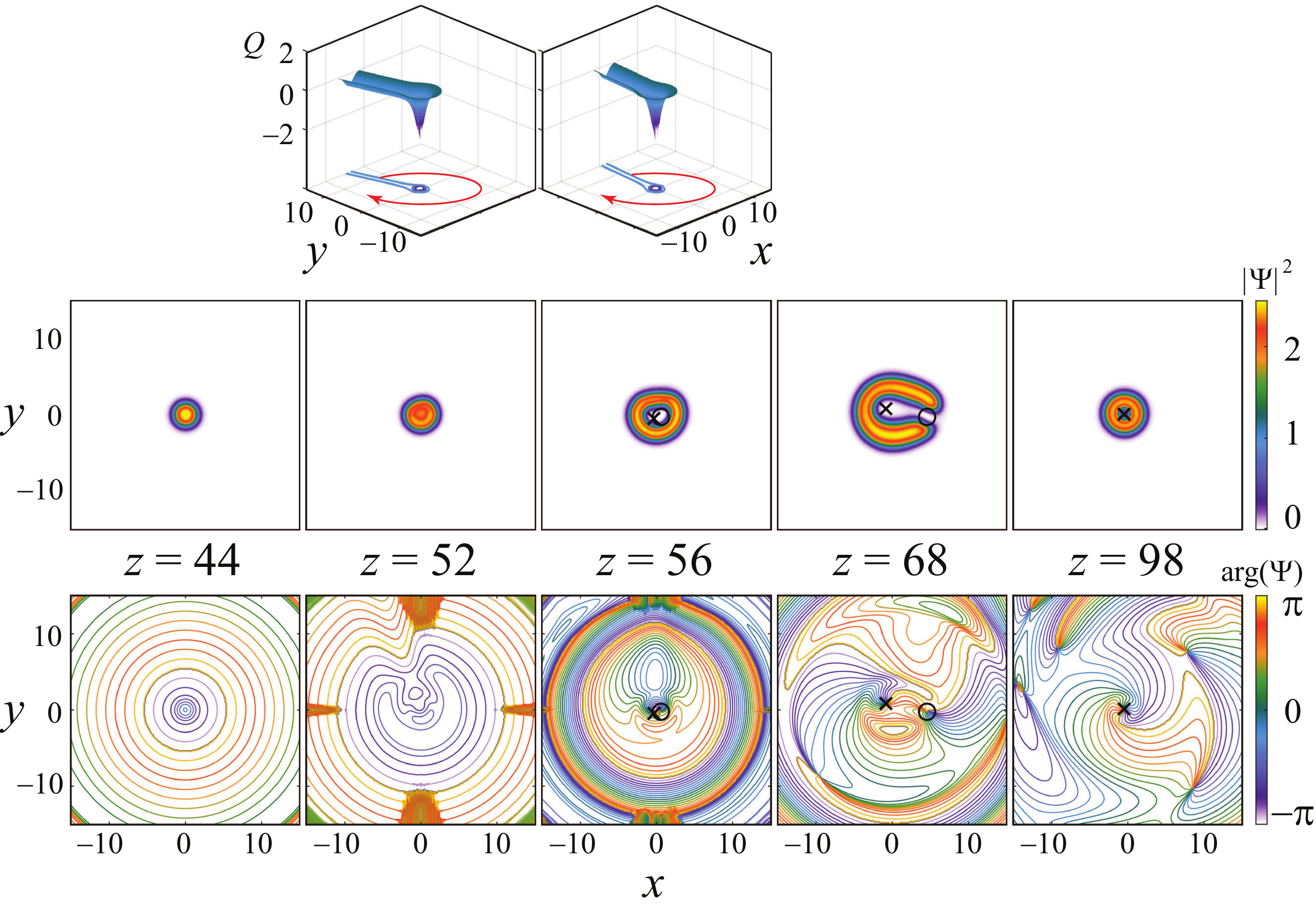}
\caption{
The same as in Fig. \ref{fig:one_to_minus_one} but for the transformation of fundamental soliton to vortex induced by repulsive central core and rotating barrier. Note that \emph{clockwise} rotation of the barrier drives the phase slip and formation of the vortex.}
\label{fig:zero_to_one_stirrer}
\end{figure}

The amplitude is chosen to be $h=0.6$, so that the potential is comparable with another terms in Eq.~\eqref{eq:CQCGLE} and the width is $r_0=2$. We use additional repelling core in the potential: $q_1=q_{c}+q_\textrm{st}$, where $q_{c}=-2\sech(r)$ produces an off-center flows at the wave beam axis and imposes the central toroidal hole. The structure of the potential implies formation of vortex-antivortex pairs near the wave beam axis. Depending on the direction of rotation of the stirrer, a vortex or an antivortex leaves the central region and moves to the periphery. Snapshots of the spatial distribution of the potential, density and phase for $\Omega=- 0.086$, $a_1=50$, $b_1=64$ are plotted in Fig.~\ref{fig:zero_to_one_stirrer}. One can estimate the speed required for vortex formation as $\Omega=L/I$, which gives the stationary vortex with charge $m=1$ when $\Omega=0.13$. In our simulations the transformation occurs for $|\Omega|\le 0.115$ in accordance with this simple estimation. Note, that it is straightforward to induce the vortex decay to fundamental soliton with $m=0$ with nonrotating barrier ($\Omega=0$).

%Note, that it is straightforward to induce the vortex decay with nonrotating barrier ($\Omega=0$). An example of decay of the vortex structure with charge $m=\pm 1$ to fundamental soliton with $m=0$ see in the Supplemental Material \cite{Suppl_Mat}.

\begin{figure}[t!]
\centering
\includegraphics[width=8.5cm]{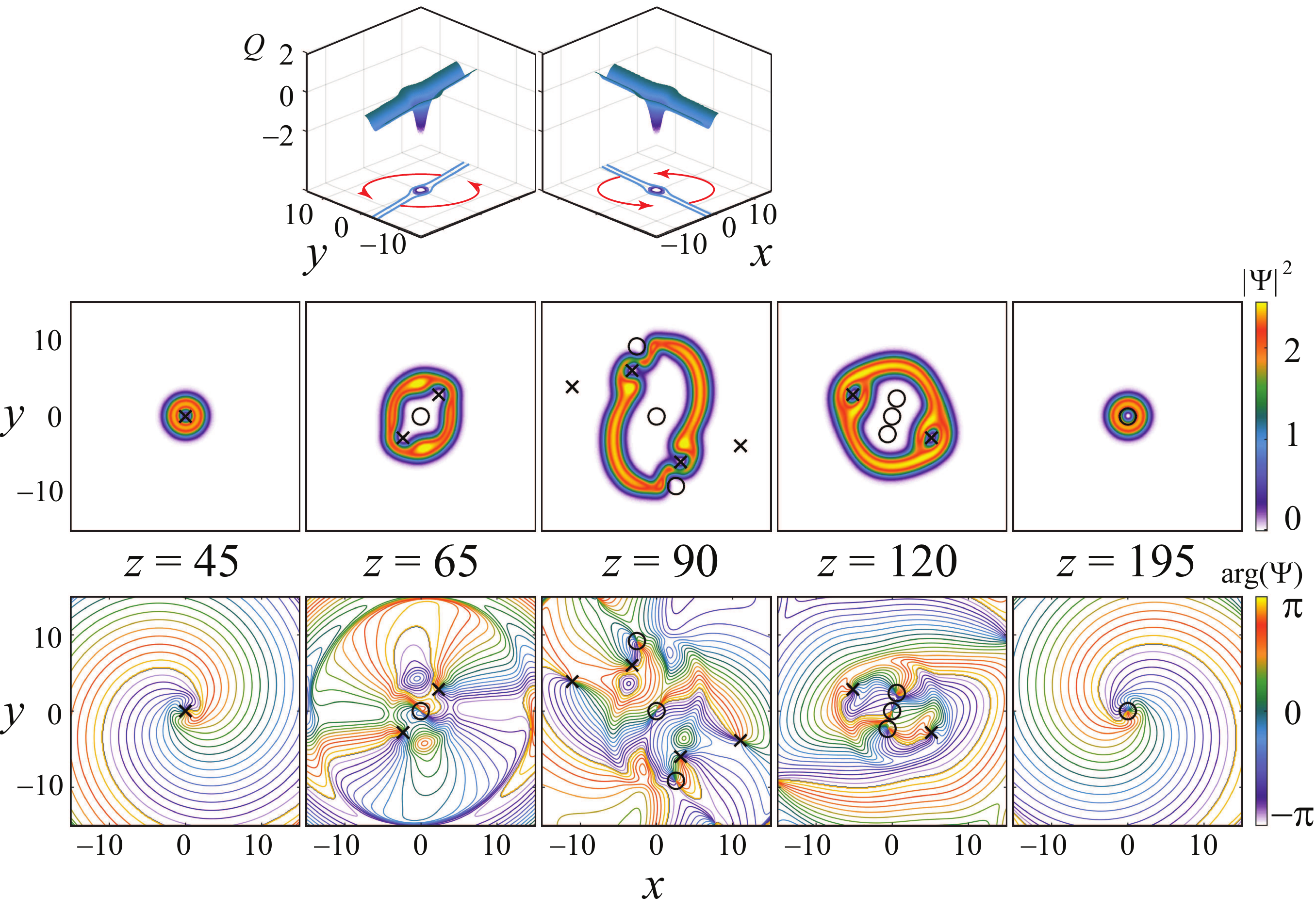}
\caption{Controllable transformation of vortex to antivortex  using rotating potential combined with repulsive core at the beam axis. Note that the \textit{anticlockwise} rotating potential change intensity and phase field distributions so that two antivortices come inside the central toroidal hole.
\label{fig:one_to_minus_one_stirrer}}
\end{figure}

Repulsive rotating potential combined with repulsive core can be used to change the vortex charge to arbitrary targeted vortex state. In particular, to transform the topological charge from $+1$ to $-1$ we use the potential of the form $q_\textrm{st}=-h \exp[-\vartheta^2(x,y,z)/r_0^2]$ with $\Omega= 0.05, ~a_1=50, ~b_1=100.$ The corresponding snapshots of the spatial distribution of the potential, density and phase for different $z$ are shown in Fig.~ \ref{fig:one_to_minus_one_stirrer}.

Thus, one can use a rotating repulsive potential to excite a vortex state from fundamental soliton in analogy with excitation of the persistent currents in atomic Bose-Einstein condensates \cite{wright2013driving, eckel2014hysteresis, yakimenko2015vortices, yakimenko2015vortex}. However, open dissipative systems with controlling potential suggest novel approaches for engineering nonlinear topological structures. The main idea is to create spatially-separated vortex-antivortex pairs and then induce a decay of part of the vortex excitations. Using this approach one can create not only single vortex
%(see the Supplemental Material \cite{Suppl_Mat})
but also generate much more complicated topological structures. To illustrate this let us discuss a transformation of fundamental soliton to a cluster of two-antivortices rotating with the relative angular momentum $L_z/E =-1.5$. For this transformation we perform four ($N_Q =  4$) control manipulations with potential \eqref{eq:Q}: (i) Firstly, the initial waveform \eqref{eq:IC} evolves to the fundamental soliton. (ii) Then at $z=50$ potentials $q_{1} = -2[\sech(r) + \sech(10-r)]$ and $q_{2} = -2[\sech(\sqrt{(x-5)^2+y^2}) + \sech(\sqrt{(x+5)^2+y^2})]$ turn on. Here term $-2\sech\left(r \right)$ creates a repulsive potential at the axis of the wave beam, and induces formation of two vortex-antivortex pairs. Terms $-2\sech\left( 10-r\right)$ and $q_{2}$ localize the pulse. (iii) After $z=77$ term $q_{2}$ is nullified and potential  $q_3=-\sech(2x)-\sech(2y)$ is applied to separate vortices and antivortices from each other. The term $q_{4}=2\sech(0.75 y-0.75 x)$ is added after $z=125$ to destroy two vortices. (iv) At $z=145$ all potentials are switched off and the resulting wave field evolves to the rotating antivortices cluster.

\begin{figure}[H]
\centering
\includegraphics[width=8.4cm]{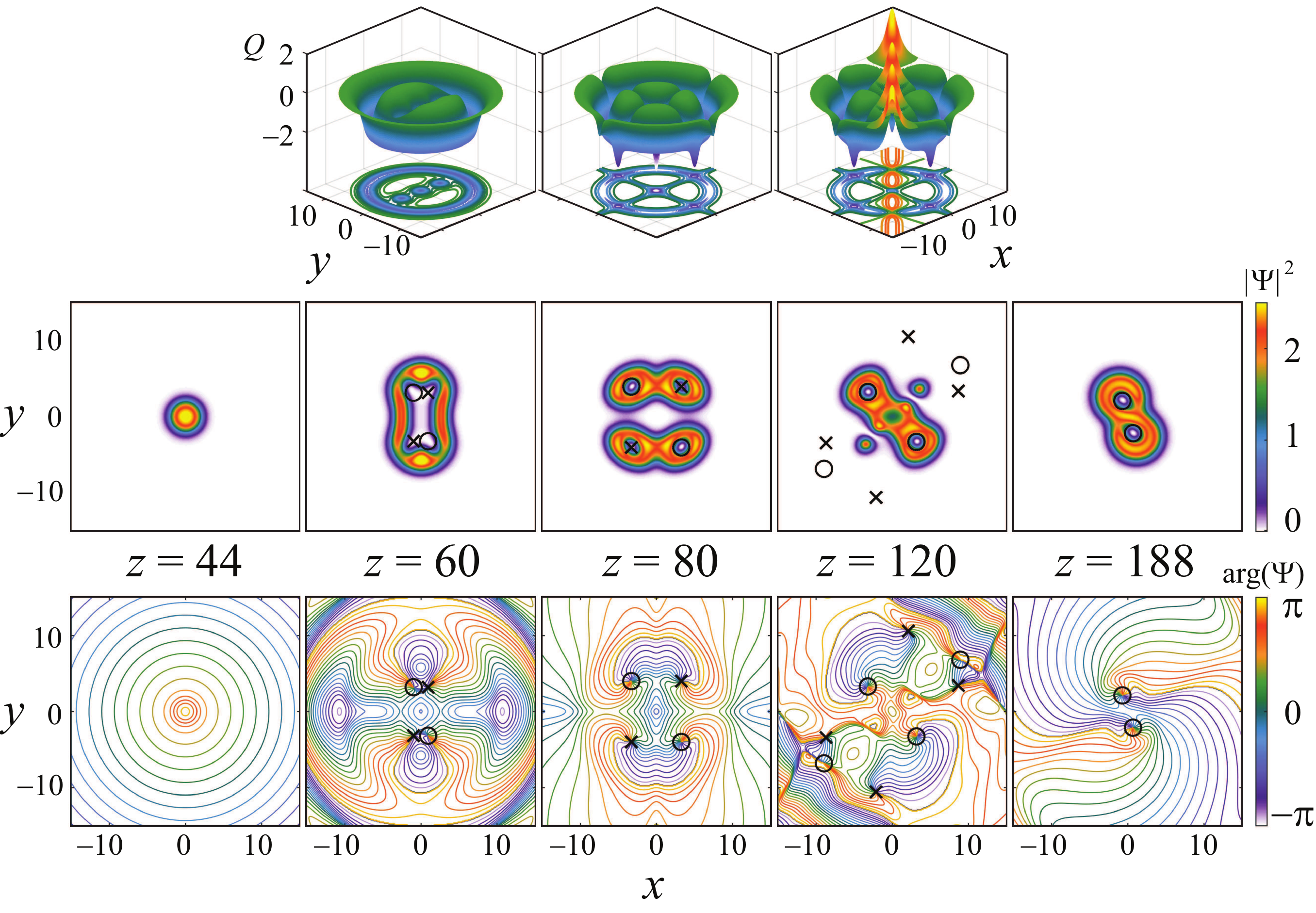}
\caption{The same as in Fig. \ref{fig:one_to_minus_one} but for the transformation of fundamental soliton to rotating cluster composed of two antivortieces. The effective topological charge of the cluster
is $L_z/E=-3/2$. In the first stage the soliton is transformed into two coupled dipoles, then the dipoles are spatially separated. Finally the attractive potential destroys the vortices, and the remaining two out-of-phase antivortices form the clockwise rotating antivortex cluster.
}\label{fig:zero_to_minus_two}
\end{figure}

Snapshots of spatial distribution of potential $Q$, intensity $|\Psi|^2$ and phase $\mbox{arg}(\Psi)$ for different $z$ are plotted in Fig.~\ref{fig:zero_to_minus_two}. This vortex cluster is a stable bound state of two out-of-phase antivortices. In practice, formation of the bound state of two vortex solitons is a challenging task since it requires fine tuning of their phase difference. Remarkable, in our case the phase tuning occurs automatically for dynamical transformation of the two pairs of two out-of-phase dipoles. We note that introducing some modifications in the described manipulations one can also release vortex and antivortex from fundamental soliton.

In summary, we analyzed dynamical transformations of different topological structures  using the model based on (2+1)D CGLE. It turns out that in a highly nonequilibrium state, driven by an external potential, various topological excitations emerge and decay. In the nonlinear media with gain and loss the energy of the vortex excitations is not conserved which permits fascinating transformations of topological structures not accessible in conservative systems.

We have analyzed spontaneous transformations of topological structures induced by external potentials. Our systematic analysis of evolution of the phase defects in the wave front opens an avenue on control over transformations of different topological structures by matching the duration of the action of the external potential during propagation of the wave beam. However, complex and irregular character of the phase transformation in the open dissipative system restricts the possibilities of control over transformation of the topological structures based on the potential duration. Furthermore generation of the vortex solitons from the fundamental solitons by simple potential is prohibited by conservation of topological charge in the system. We  found a series of external potentials that drive completely controllable transitions of the dissipative soliton or vortex into the structures with required topological charge. Moreover, using the method developed in this work, we demonstrate formation of complex solitonic and vortex structures. The proposed method of controllable transformation of topological structures may open new prospects for future applications in optical signal processing.

%Acknowledgments
The authors are grateful for a support from Jilin University, China.

%\bibliography{Soliton}

\begin{thebibliography}{99}
\makeatletter
\providecommand \@ifxundefined [1]{%
 \@ifx{#1\undefined}
}%
\providecommand \@ifnum [1]{%
 \ifnum #1\expandafter \@firstoftwo
 \else \expandafter \@secondoftwo
 \fi
}%
\providecommand \@ifx [1]{%
 \ifx #1\expandafter \@firstoftwo
 \else \expandafter \@secondoftwo
 \fi
}%
\providecommand \natexlab [1]{#1}%
\providecommand \enquote  [1]{``#1''}%
\providecommand \bibnamefont  [1]{#1}%
\providecommand \bibfnamefont [1]{#1}%
\providecommand \citenamefont [1]{#1}%
\providecommand \href@noop [0]{\@secondoftwo}%
\providecommand \href [0]{\begingroup \@sanitize@url \@href}%
\providecommand \@href[1]{\@@startlink{#1}\@@href}%
\providecommand \@@href[1]{\endgroup#1\@@endlink}%
\providecommand \@sanitize@url [0]{\catcode `\\12\catcode `\$12\catcode
  `\&12\catcode `\#12\catcode `\^12\catcode `\_12\catcode `\%12\relax}%
\providecommand \@@startlink[1]{}%
\providecommand \@@endlink[0]{}%
\providecommand \url  [0]{\begingroup\@sanitize@url \@url }%
\providecommand \@url [1]{\endgroup\@href {#1}{\urlprefix }}%
\providecommand \urlprefix  [0]{URL }%
\providecommand \Eprint [0]{\href }%
\providecommand \doibase [0]{http://dx.doi.org/}%
\providecommand \selectlanguage [0]{\@gobble}%
\providecommand \bibinfo  [0]{\@secondoftwo}%
\providecommand \bibfield  [0]{\@secondoftwo}%
\providecommand \translation [1]{[#1]}%
\providecommand \BibitemOpen [0]{}%
\providecommand \bibitemStop [0]{}%
\providecommand \bibitemNoStop [0]{.\EOS\space}%
\providecommand \EOS [0]{\spacefactor3000\relax}%
\providecommand \BibitemShut  [1]{\csname bibitem#1\endcsname}%
\let\auto@bib@innerbib\@empty
%</preamble>
\bibitem [{\citenamefont {Pismen}(1999)}]{Pismen_Book}%
  \BibitemOpen
  \bibfield  {author} {\bibinfo {author} {\bibfnamefont {L.~M.}\ \bibnamefont
  {Pismen}},\ }\href@noop {} {\emph {\bibinfo {title} {Vortices in Nonlinear
  Fields}}}\ (\bibinfo  {publisher} {Clarendon Press},\ \bibinfo {address}
  {Oxford},\ \bibinfo {year} {1999})\BibitemShut {NoStop}%
\bibitem [{\citenamefont {Desyatnikov}\ \emph {et~al.}(2005)\citenamefont
  {Desyatnikov}, \citenamefont {Kivshar},\ and\ \citenamefont
  {Torner}}]{Desyatnikov_PO_2005}%
  \BibitemOpen
  \bibfield  {author} {\bibinfo {author} {\bibfnamefont {A.~S.}\ \bibnamefont
  {Desyatnikov}}, \bibinfo {author} {\bibfnamefont {Y.}~\bibnamefont
  {Kivshar}}, \ and\ \bibinfo {author} {\bibfnamefont {L.}~\bibnamefont
  {Torner}},\ }\bibfield  {title} {\enquote {\bibinfo {title} {Optical vortices
  and vortex solitons},}\ }in\ \href@noop {} {\emph {\bibinfo {booktitle}
  {Progress in Optics}}},\ Vol.~\bibinfo {volume} {47},\ \bibinfo {editor}
  {edited by\ \bibinfo {editor} {\bibfnamefont {E.}~\bibnamefont {Wolf}}}\
  (\bibinfo  {publisher} {Elsevier},\ \bibinfo {address} {The Netherlands},\
  \bibinfo {year} {2005})\ Chap.~\bibinfo {chapter} {5}, pp.\ \bibinfo {pages}
  {291--391}\BibitemShut {NoStop}%
\bibitem [{\citenamefont {Soskin}\ \emph {et~al.}(2017)\citenamefont {Soskin},
  \citenamefont {Boriskina}, \citenamefont {Chong}, \citenamefont {Dennis},\
  and\ \citenamefont {Desyatnikov}}]{Soskin_JO_2017}%
  \BibitemOpen
  \bibfield  {author} {\bibinfo {author} {\bibfnamefont {M.}~\bibnamefont
  {Soskin}}, \bibinfo {author} {\bibfnamefont {S.~V.}\ \bibnamefont
  {Boriskina}}, \bibinfo {author} {\bibfnamefont {Y.}~\bibnamefont {Chong}},
  \bibinfo {author} {\bibfnamefont {M.~R.}\ \bibnamefont {Dennis}}, \ and\
  \bibinfo {author} {\bibfnamefont {A.}~\bibnamefont {Desyatnikov}},\
  }\bibfield  {title} {\enquote {\bibinfo {title} {Singular optics and
  topological photonics},}\ }\href
  {http://stacks.iop.org/2040-8986/19/i=1/a=010401} {\bibfield  {journal}
  {\bibinfo  {journal} {J. Opt.}\ }\textbf {\bibinfo {volume} {19}},\ \bibinfo
  {pages} {010401} (\bibinfo {year} {2017})}\BibitemShut {NoStop}%
\bibitem [{\citenamefont {Lagoudakis}\ \emph {et~al.}(2008)\citenamefont
  {Lagoudakis}, \citenamefont {Wouters}, \citenamefont {Richard}, \citenamefont
  {Baas}, \citenamefont {Carusotto}, \citenamefont {Andr\'e}, \citenamefont
  {Dang},\ and\ \citenamefont {Deveaud-Pl\'edran}}]{Lagoudakis_NPh_2008}%
  \BibitemOpen
  \bibfield  {author} {\bibinfo {author} {\bibfnamefont {K.~G.}\ \bibnamefont
  {Lagoudakis}}, \bibinfo {author} {\bibfnamefont {M.}~\bibnamefont {Wouters}},
  \bibinfo {author} {\bibfnamefont {M.}~\bibnamefont {Richard}}, \bibinfo
  {author} {\bibfnamefont {A.}~\bibnamefont {Baas}}, \bibinfo {author}
  {\bibfnamefont {I.}~\bibnamefont {Carusotto}}, \bibinfo {author}
  {\bibfnamefont {R.}~\bibnamefont {Andr\'e}}, \bibinfo {author} {\bibfnamefont
  {Le~Si}\ \bibnamefont {Dang}}, \ and\ \bibinfo {author} {\bibfnamefont
  {B.}~\bibnamefont {Deveaud-Pl\'edran}},\ }\bibfield  {title} {\enquote
  {\bibinfo {title} {Quantized vortices in an exciton-polariton condensate},}\
  }\href {\doibase 10.1038/nphys1051} {\bibfield  {journal} {\bibinfo
  {journal} {Nat. Physics}\ }\textbf {\bibinfo {volume} {4}},\ \bibinfo {pages}
  {706} (\bibinfo {year} {2008})}\BibitemShut {NoStop}%
\bibitem [{\citenamefont {Lagoudakis}\ \emph {et~al.}(2009)\citenamefont
  {Lagoudakis}, \citenamefont {Ostatnick{\'y}}, \citenamefont {Kavokin},
  \citenamefont {Rubo}, \citenamefont {Andr{\'e}},\ and\ \citenamefont
  {Deveaud-Pl{\'e}dran}}]{Lagoudakis_S_2009}%
  \BibitemOpen
  \bibfield  {author} {\bibinfo {author} {\bibfnamefont {K.~G.}\ \bibnamefont
  {Lagoudakis}}, \bibinfo {author} {\bibfnamefont {T.}~\bibnamefont
  {Ostatnick{\'y}}}, \bibinfo {author} {\bibfnamefont {A.~V.}\ \bibnamefont
  {Kavokin}}, \bibinfo {author} {\bibfnamefont {Y.~G.}\ \bibnamefont {Rubo}},
  \bibinfo {author} {\bibfnamefont {R.}~\bibnamefont {Andr{\'e}}}, \ and\
  \bibinfo {author} {\bibfnamefont {B.}~\bibnamefont {Deveaud-Pl{\'e}dran}},\
  }\bibfield  {title} {\enquote {\bibinfo {title} {Observation of half-quantum
  vortices in an exciton-polariton condensate},}\ }\href {\doibase
  10.1126/science.1177980} {\bibfield  {journal} {\bibinfo  {journal}
  {Science}\ }\textbf {\bibinfo {volume} {326}},\ \bibinfo {pages} {974--976}
  (\bibinfo {year} {2009})}\BibitemShut {NoStop}%
\bibitem [{\citenamefont {Ostrovskaya}\ \emph {et~al.}(2012)\citenamefont
  {Ostrovskaya}, \citenamefont {Abdullaev}, \citenamefont {Desyatnikov},
  \citenamefont {Fraser},\ and\ \citenamefont
  {Kivshar}}]{Ostrovskaya_PRA_2012}%
  \BibitemOpen
  \bibfield  {author} {\bibinfo {author} {\bibfnamefont {E.~A.}\ \bibnamefont
  {Ostrovskaya}}, \bibinfo {author} {\bibfnamefont {J.}~\bibnamefont
  {Abdullaev}}, \bibinfo {author} {\bibfnamefont {A.~S.}\ \bibnamefont
  {Desyatnikov}}, \bibinfo {author} {\bibfnamefont {M.~D.}\ \bibnamefont
  {Fraser}}, \ and\ \bibinfo {author} {\bibfnamefont {Y.~S.}\ \bibnamefont
  {Kivshar}},\ }\bibfield  {title} {\enquote {\bibinfo {title} {Dissipative
  solitons and vortices in polariton {Bose-Einstein} condensates},}\ }\href
  {\doibase 10.1103/PhysRevA.86.013636} {\bibfield  {journal} {\bibinfo
  {journal} {Phys. Rev. A}\ }\textbf {\bibinfo {volume} {86}},\ \bibinfo
  {pages} {013636} (\bibinfo {year} {2012})}\BibitemShut {NoStop}%
\bibitem [{\citenamefont {Dall}\ \emph {et~al.}(2014)\citenamefont {Dall},
  \citenamefont {Fraser}, \citenamefont {Desyatnikov}, \citenamefont {Li},
  \citenamefont {Brodbeck}, \citenamefont {Kamp}, \citenamefont {Schneider},
  \citenamefont {H\"ofling},\ and\ \citenamefont
  {Ostrovskaya}}]{Dall_PRL_2014}%
  \BibitemOpen
  \bibfield  {author} {\bibinfo {author} {\bibfnamefont {R.}~\bibnamefont
  {Dall}}, \bibinfo {author} {\bibfnamefont {M.~D.}\ \bibnamefont {Fraser}},
  \bibinfo {author} {\bibfnamefont {A.~S.}\ \bibnamefont {Desyatnikov}},
  \bibinfo {author} {\bibfnamefont {G.}~\bibnamefont {Li}}, \bibinfo {author}
  {\bibfnamefont {S.}~\bibnamefont {Brodbeck}}, \bibinfo {author}
  {\bibfnamefont {M.}~\bibnamefont {Kamp}}, \bibinfo {author} {\bibfnamefont
  {C.}~\bibnamefont {Schneider}}, \bibinfo {author} {\bibfnamefont
  {S.}~\bibnamefont {H\"ofling}}, \ and\ \bibinfo {author} {\bibfnamefont
  {E.~A.}\ \bibnamefont {Ostrovskaya}},\ }\bibfield  {title} {\enquote
  {\bibinfo {title} {{Creation of Orbital Angular Momentum States with Chiral
  Polaritonic Lenses}},}\ }\href {\doibase 10.1103/PhysRevLett.113.200404}
  {\bibfield  {journal} {\bibinfo  {journal} {Phys. Rev. Lett.}\ }\textbf
  {\bibinfo {volume} {113}},\ \bibinfo {pages} {200404} (\bibinfo {year}
  {2014})}\BibitemShut {NoStop}%
\bibitem [{\citenamefont {Ma}\ \emph {et~al.}(2017)\citenamefont {Ma},
  \citenamefont {Egorov},\ and\ \citenamefont {Schumacher}}]{Ma_PRL_2017}%
  \BibitemOpen
  \bibfield  {author} {\bibinfo {author} {\bibfnamefont {X.}~\bibnamefont
  {Ma}}, \bibinfo {author} {\bibfnamefont {O.~A.}\ \bibnamefont {Egorov}}, \
  and\ \bibinfo {author} {\bibfnamefont {S.}~\bibnamefont {Schumacher}},\
  }\bibfield  {title} {\enquote {\bibinfo {title} {{Creation and Manipulation
  of Stable Dark Solitons and Vortices in Microcavity Polariton
  Condensates}},}\ }\href {\doibase 10.1103/PhysRevLett.118.157401} {\bibfield
  {journal} {\bibinfo  {journal} {Phys. Rev. Lett.}\ }\textbf {\bibinfo
  {volume} {118}},\ \bibinfo {pages} {157401} (\bibinfo {year}
  {2017})}\BibitemShut {NoStop}%
\bibitem [{\citenamefont {Sigurdsson}\ \emph {et~al.}(2014)\citenamefont
  {Sigurdsson}, \citenamefont {Egorov}, \citenamefont {Ma}, \citenamefont
  {Shelykh},\ and\ \citenamefont {Liew}}]{Sigurdsson_PRB_2014}%
  \BibitemOpen
  \bibfield  {author} {\bibinfo {author} {\bibfnamefont {H.}~\bibnamefont
  {Sigurdsson}}, \bibinfo {author} {\bibfnamefont {O.~A.}\ \bibnamefont
  {Egorov}}, \bibinfo {author} {\bibfnamefont {X.}~\bibnamefont {Ma}}, \bibinfo
  {author} {\bibfnamefont {I.~A.}\ \bibnamefont {Shelykh}}, \ and\ \bibinfo
  {author} {\bibfnamefont {T.~C.~H.}\ \bibnamefont {Liew}},\ }\bibfield
  {title} {\enquote {\bibinfo {title} {Information processing with
  topologically protected vortex memories in exciton-polariton condensates},}\
  }\href {\doibase 10.1103/PhysRevB.90.014504} {\bibfield  {journal} {\bibinfo
  {journal} {Phys. Rev. B}\ }\textbf {\bibinfo {volume} {90}},\ \bibinfo
  {pages} {014504} (\bibinfo {year} {2014})}\BibitemShut {NoStop}%
\bibitem [{\citenamefont {Ma}\ \emph {et~al.}(2016)\citenamefont {Ma},
  \citenamefont {Peschel},\ and\ \citenamefont {Egorov}}]{Ma_PRB_2016}%
  \BibitemOpen
  \bibfield  {author} {\bibinfo {author} {\bibfnamefont {X.}~\bibnamefont
  {Ma}}, \bibinfo {author} {\bibfnamefont {U.}~\bibnamefont {Peschel}}, \ and\
  \bibinfo {author} {\bibfnamefont {O.~A.}\ \bibnamefont {Egorov}},\ }\bibfield
   {title} {\enquote {\bibinfo {title} {Incoherent control of topological
  charges in nonequilibrium polariton condensates},}\ }\href {\doibase
  10.1103/PhysRevB.93.035315} {\bibfield  {journal} {\bibinfo  {journal} {Phys.
  Rev. B}\ }\textbf {\bibinfo {volume} {93}},\ \bibinfo {pages} {035315}
  (\bibinfo {year} {2016})}\BibitemShut {NoStop}%
\bibitem [{\citenamefont {Ma}\ and\ \citenamefont
  {Schumacher}(2018)}]{Ma_PRL_2018}%
  \BibitemOpen
  \bibfield  {author} {\bibinfo {author} {\bibfnamefont {X.}~\bibnamefont
  {Ma}}\ and\ \bibinfo {author} {\bibfnamefont {S.}~\bibnamefont
  {Schumacher}},\ }\bibfield  {title} {\enquote {\bibinfo {title} {{Vortex
  Multistability and {Bessel} Vortices in Polariton Condensates}},}\ }\href
  {\doibase 10.1103/PhysRevLett.121.227404} {\bibfield  {journal} {\bibinfo
  {journal} {Phys. Rev. Lett.}\ }\textbf {\bibinfo {volume} {121}},\ \bibinfo
  {pages} {227404} (\bibinfo {year} {2018})}\BibitemShut {NoStop}%
\bibitem [{\citenamefont {Fraser}\ \emph {et~al.}(2009)\citenamefont {Fraser},
  \citenamefont {Roumpos},\ and\ \citenamefont {Yamamoto}}]{Fraser_NJP_2009}%
  \BibitemOpen
  \bibfield  {author} {\bibinfo {author} {\bibfnamefont {M.~D.}\ \bibnamefont
  {Fraser}}, \bibinfo {author} {\bibfnamefont {G.}~\bibnamefont {Roumpos}}, \
  and\ \bibinfo {author} {\bibfnamefont {Y.}~\bibnamefont {Yamamoto}},\
  }\bibfield  {title} {\enquote {\bibinfo {title} {Vortex-antivortex pair
  dynamics in an exciton-polariton condensate},}\ }\href {\doibase
  10.1088/1367-2630/11/11/113048} {\bibfield  {journal} {\bibinfo  {journal}
  {New J. Phys.}\ }\textbf {\bibinfo {volume} {11}},\ \bibinfo {pages} {113048}
  (\bibinfo {year} {2009})}\BibitemShut {NoStop}%
\bibitem [{\citenamefont {Boulier}\ \emph {et~al.}(2016)\citenamefont
  {Boulier}, \citenamefont {Cancellieri}, \citenamefont {Sangouard},
  \citenamefont {Hivet}, \citenamefont {Glorieux}, \citenamefont {Giacobino},\
  and\ \citenamefont {Bramati}}]{Boulier_CRP_2016}%
  \BibitemOpen
  \bibfield  {author} {\bibinfo {author} {\bibfnamefont {T.}~\bibnamefont
  {Boulier}}, \bibinfo {author} {\bibfnamefont {E.}~\bibnamefont
  {Cancellieri}}, \bibinfo {author} {\bibfnamefont {N.~D.}\ \bibnamefont
  {Sangouard}}, \bibinfo {author} {\bibfnamefont {R.}~\bibnamefont {Hivet}},
  \bibinfo {author} {\bibfnamefont {Q.}~\bibnamefont {Glorieux}}, \bibinfo
  {author} {\bibfnamefont {E.}~\bibnamefont {Giacobino}}, \ and\ \bibinfo
  {author} {\bibfnamefont {A.}~\bibnamefont {Bramati}},\ }\bibfield  {title}
  {\enquote {\bibinfo {title} {Lattices of quantized vortices in polariton
  superfluids},}\ }\href {\doibase https://doi.org/10.1016/j.crhy.2016.05.005}
  {\bibfield  {journal} {\bibinfo  {journal} {C. R. Phys.}\ }\textbf {\bibinfo
  {volume} {17}},\ \bibinfo {pages} {893--907} (\bibinfo {year}
  {2016})}\BibitemShut {NoStop}%
\bibitem [{\citenamefont {Ma}\ and\ \citenamefont
  {Schumacher}(2017)}]{Ma_PRB_2017}%
  \BibitemOpen
  \bibfield  {author} {\bibinfo {author} {\bibfnamefont {X.}~\bibnamefont
  {Ma}}\ and\ \bibinfo {author} {\bibfnamefont {S.}~\bibnamefont
  {Schumacher}},\ }\bibfield  {title} {\enquote {\bibinfo {title}
  {Vortex-vortex control in exciton-polariton condensates},}\ }\href {\doibase
  10.1103/PhysRevB.95.235301} {\bibfield  {journal} {\bibinfo  {journal} {Phys.
  Rev. B}\ }\textbf {\bibinfo {volume} {95}},\ \bibinfo {pages} {235301}
  (\bibinfo {year} {2017})}\BibitemShut {NoStop}%
\bibitem [{\citenamefont {Cross}\ and\ \citenamefont
  {Hohenberg}(1993)}]{Cross_RMP_1993}%
  \BibitemOpen
  \bibfield  {author} {\bibinfo {author} {\bibfnamefont {M.~C.}\ \bibnamefont
  {Cross}}\ and\ \bibinfo {author} {\bibfnamefont {P.~C.}\ \bibnamefont
  {Hohenberg}},\ }\bibfield  {title} {\enquote {\bibinfo {title} {{Pattern
  Formation Outside of Equilibrium}},}\ }\href {\doibase
  10.1103/RevModPhys.65.851} {\bibfield  {journal} {\bibinfo  {journal} {Rev.
  Mod. Phys.}\ }\textbf {\bibinfo {volume} {65}},\ \bibinfo {pages} {851--1112}
  (\bibinfo {year} {1993})}\BibitemShut {NoStop}%
\bibitem [{\citenamefont {Aranson}\ and\ \citenamefont
  {Kramer}(2002)}]{Aranson_RMP_2002}%
  \BibitemOpen
  \bibfield  {author} {\bibinfo {author} {\bibfnamefont {I.~S.}\ \bibnamefont
  {Aranson}}\ and\ \bibinfo {author} {\bibfnamefont {L.}~\bibnamefont
  {Kramer}},\ }\bibfield  {title} {\enquote {\bibinfo {title} {The world of the
  complex {Ginzburg-Landau} equation},}\ }\href {\doibase
  10.1103/RevModPhys.74.99} {\bibfield  {journal} {\bibinfo  {journal} {Rev.
  Mod. Phys.}\ }\textbf {\bibinfo {volume} {74}},\ \bibinfo {pages} {99--143}
  (\bibinfo {year} {2002})}\BibitemShut {NoStop}%
\bibitem [{\citenamefont {Akhmediev}\ and\ \citenamefont {{A. Ankiewicz
  (Eds.)}}(2005)}]{Akhmediev_Book1}%
  \BibitemOpen
  \bibfield  {author} {\bibinfo {author} {\bibfnamefont {N.}~\bibnamefont
  {Akhmediev}}\ and\ \bibinfo {author} {\bibnamefont {{A. Ankiewicz (Eds.)}}},\
  }\href@noop {} {\emph {\bibinfo {title} {Dissipative Solitons}}}\ (\bibinfo
  {publisher} {Springer},\ \bibinfo {address} {Berlin},\ \bibinfo {year}
  {2005})\BibitemShut {NoStop}%
\bibitem [{\citenamefont {{N. Akhmediev and {A. Ankiewicz
  (Eds.)}}}(2008)}]{Akhmediev_Book2}%
  \BibitemOpen
  \bibfield  {author} {\bibinfo {author} {\bibnamefont {{N. Akhmediev and {A.
  Ankiewicz (Eds.)}}}},\ }\href@noop {} {\emph {\bibinfo {title} {Dissipative
  Solitons: From Optics to Biology and Medicine}}}\ (\bibinfo  {publisher}
  {Springer},\ \bibinfo {address} {Berlin},\ \bibinfo {year}
  {2008})\BibitemShut {NoStop}%
\bibitem [{\citenamefont {Garc\'ia-Morales}\ and\ \citenamefont
  {Krischer}(2012)}]{Garcia-Morales_CP_2012}%
  \BibitemOpen
  \bibfield  {author} {\bibinfo {author} {\bibfnamefont {V.}~\bibnamefont
  {Garc\'ia-Morales}}\ and\ \bibinfo {author} {\bibfnamefont {K.}~\bibnamefont
  {Krischer}},\ }\bibfield  {title} {\enquote {\bibinfo {title} {{The Complex
  {Ginzburg-Landau} Equation: An Introduction}},}\ }\href@noop {} {\bibfield
  {journal} {\bibinfo  {journal} {Contemp. Phys.}\ }\textbf {\bibinfo {volume}
  {53}},\ \bibinfo {pages} {79--95} (\bibinfo {year} {2012})}\BibitemShut
  {NoStop}%
\bibitem [{\citenamefont {Liehr}(2013)}]{Liehr_Book}%
  \BibitemOpen
  \bibfield  {author} {\bibinfo {author} {\bibfnamefont {A.}~\bibnamefont
  {Liehr}},\ }\href@noop {} {\emph {\bibinfo {title} {Dissipative Solitons in
  Reaction Diffusion Systems}}}\ (\bibinfo  {publisher} {Springer},\ \bibinfo
  {address} {Berlin},\ \bibinfo {year} {2013})\BibitemShut {NoStop}%
\bibitem [{\citenamefont {Boardman}\ \emph {et~al.}(2005)\citenamefont
  {Boardman}, \citenamefont {Velasco},\ and\ \citenamefont
  {Egan}}]{Boardman_Chapter_2005}%
  \BibitemOpen
  \bibfield  {author} {\bibinfo {author} {\bibfnamefont {A.D.}\ \bibnamefont
  {Boardman}}, \bibinfo {author} {\bibfnamefont {L.}~\bibnamefont {Velasco}}, \
  and\ \bibinfo {author} {\bibfnamefont {P.}~\bibnamefont {Egan}},\ }\bibfield
  {title} {\enquote {\bibinfo {title} {Dissipative magneto-optic solitons},}\
  }in\ \href@noop {} {\emph {\bibinfo {booktitle} {Dissipative Solitons}}},\
  \bibinfo {editor} {edited by\ \bibinfo {editor} {\bibfnamefont
  {N.}~\bibnamefont {Akhmediev}}\ and\ \bibinfo {editor} {\bibfnamefont
  {A.}~\bibnamefont {Ankiewicz}}}\ (\bibinfo  {publisher} {Springer},\ \bibinfo
  {address} {Berlin},\ \bibinfo {year} {2005})\ Chap.~\bibinfo {chapter} {2},
  pp.\ \bibinfo {pages} {19--37}\BibitemShut {NoStop}%
\bibitem [{\citenamefont {Boardman}\ and\ \citenamefont
  {Velasco}(2006)}]{Boardman_2006}%
  \BibitemOpen
  \bibfield  {author} {\bibinfo {author} {\bibfnamefont {A.~D.}\ \bibnamefont
  {Boardman}}\ and\ \bibinfo {author} {\bibfnamefont {L.}~\bibnamefont
  {Velasco}},\ }\bibfield  {title} {\enquote {\bibinfo {title} {Gyroelectric
  cubic-quintic dissipative solitons},}\ }\href {\doibase
  10.1109/JSTQE.2006.872718} {\bibfield  {journal} {\bibinfo  {journal} {IEEE
  J. Sel. Top. Quantum Electron.}\ }\textbf {\bibinfo {volume} {12}},\ \bibinfo
  {pages} {388--397} (\bibinfo {year} {2006})}\BibitemShut {NoStop}%
\bibitem [{\citenamefont {Kochetov}\ \emph {et~al.}(2017)\citenamefont
  {Kochetov}, \citenamefont {Vasylieva}, \citenamefont {Kochetova},
  \citenamefont {Sun},\ and\ \citenamefont {Tuz}}]{OptLett_2017}%
  \BibitemOpen
  \bibfield  {author} {\bibinfo {author} {\bibfnamefont {B.~A.}\ \bibnamefont
  {Kochetov}}, \bibinfo {author} {\bibfnamefont {I.}~\bibnamefont {Vasylieva}},
  \bibinfo {author} {\bibfnamefont {L.~A.}\ \bibnamefont {Kochetova}}, \bibinfo
  {author} {\bibfnamefont {H.-B.}\ \bibnamefont {Sun}}, \ and\ \bibinfo
  {author} {\bibfnamefont {V.~R.}\ \bibnamefont {Tuz}},\ }\bibfield  {title}
  {\enquote {\bibinfo {title} {Control of dissipative solitons in a
  magneto-optic planar waveguide},}\ }\href {\doibase 10.1364/OL.42.000531}
  {\bibfield  {journal} {\bibinfo  {journal} {Opt. Lett.}\ }\textbf {\bibinfo
  {volume} {42}},\ \bibinfo {pages} {531--534} (\bibinfo {year}
  {2017})}\BibitemShut {NoStop}%
\bibitem [{\citenamefont {He}\ \emph {et~al.}(2009)\citenamefont {He},
  \citenamefont {Malomed}, \citenamefont {Mihalache}, \citenamefont {Liu},
  \citenamefont {Huang}, \citenamefont {Yang},\ and\ \citenamefont
  {Wang}}]{He_OL_2009}%
  \BibitemOpen
  \bibfield  {author} {\bibinfo {author} {\bibfnamefont {Y.~J.}\ \bibnamefont
  {He}}, \bibinfo {author} {\bibfnamefont {B.~A.}\ \bibnamefont {Malomed}},
  \bibinfo {author} {\bibfnamefont {D.}~\bibnamefont {Mihalache}}, \bibinfo
  {author} {\bibfnamefont {B.}~\bibnamefont {Liu}}, \bibinfo {author}
  {\bibfnamefont {H.~C.}\ \bibnamefont {Huang}}, \bibinfo {author}
  {\bibfnamefont {H.}~\bibnamefont {Yang}}, \ and\ \bibinfo {author}
  {\bibfnamefont {H.~Z.}\ \bibnamefont {Wang}},\ }\bibfield  {title} {\enquote
  {\bibinfo {title} {Bound states of one-, two-, and three-dimensional solitons
  in complex {Ginzburg--Landau} equations with a linear potential},}\ }\href
  {\doibase 10.1364/OL.34.002976} {\bibfield  {journal} {\bibinfo  {journal}
  {Opt. Lett.}\ }\textbf {\bibinfo {volume} {34}},\ \bibinfo {pages}
  {2976--2978} (\bibinfo {year} {2009})}\BibitemShut {NoStop}%
\bibitem [{\citenamefont {Leblond}\ \emph {et~al.}(2009)\citenamefont
  {Leblond}, \citenamefont {Malomed},\ and\ \citenamefont
  {Mihalache}}]{Leblond_PRA_2009}%
  \BibitemOpen
  \bibfield  {author} {\bibinfo {author} {\bibfnamefont {H.}~\bibnamefont
  {Leblond}}, \bibinfo {author} {\bibfnamefont {B.~A.}\ \bibnamefont
  {Malomed}}, \ and\ \bibinfo {author} {\bibfnamefont {D.}~\bibnamefont
  {Mihalache}},\ }\bibfield  {title} {\enquote {\bibinfo {title} {Stable vortex
  solitons in the {Ginzburg-Landau} model of a two-dimensional lasing medium
  with a transverse grating},}\ }\href {\doibase 10.1103/PhysRevA.80.033835}
  {\bibfield  {journal} {\bibinfo  {journal} {Phys. Rev. A}\ }\textbf {\bibinfo
  {volume} {80}},\ \bibinfo {pages} {033835} (\bibinfo {year}
  {2009})}\BibitemShut {NoStop}%
\bibitem [{\citenamefont {Liu}\ \emph {et~al.}(2010)\citenamefont {Liu},
  \citenamefont {He}, \citenamefont {Malomed}, \citenamefont {Wang},
  \citenamefont {Kevrekidis}, \citenamefont {Wang}, \citenamefont {Leng},
  \citenamefont {Qiu},\ and\ \citenamefont {Wang}}]{Liu_OL_2010}%
  \BibitemOpen
  \bibfield  {author} {\bibinfo {author} {\bibfnamefont {B.}~\bibnamefont
  {Liu}}, \bibinfo {author} {\bibfnamefont {Y.-J.}\ \bibnamefont {He}},
  \bibinfo {author} {\bibfnamefont {B.~A.}\ \bibnamefont {Malomed}}, \bibinfo
  {author} {\bibfnamefont {X.-S.}\ \bibnamefont {Wang}}, \bibinfo {author}
  {\bibfnamefont {P.~G.}\ \bibnamefont {Kevrekidis}}, \bibinfo {author}
  {\bibfnamefont {T.-B.}\ \bibnamefont {Wang}}, \bibinfo {author}
  {\bibfnamefont {F.-C.}\ \bibnamefont {Leng}}, \bibinfo {author}
  {\bibfnamefont {Z.-R.}\ \bibnamefont {Qiu}}, \ and\ \bibinfo {author}
  {\bibfnamefont {H.-Z.}\ \bibnamefont {Wang}},\ }\bibfield  {title} {\enquote
  {\bibinfo {title} {Continuous generation of soliton patterns in
  two-dimensional dissipative media by razor, dagger, and needle potentials},}\
  }\href {\doibase 10.1364/OL.35.001974} {\bibfield  {journal} {\bibinfo
  {journal} {Opt. Lett.}\ }\textbf {\bibinfo {volume} {35}},\ \bibinfo {pages}
  {1974--1976} (\bibinfo {year} {2010})}\BibitemShut {NoStop}%
\bibitem [{\citenamefont {He}\ \emph {et~al.}(2010)\citenamefont {He},
  \citenamefont {Malomed}, \citenamefont {Ye},\ and\ \citenamefont
  {Hu}}]{He_JOSAB_2010}%
  \BibitemOpen
  \bibfield  {author} {\bibinfo {author} {\bibfnamefont {Y.-J.}\ \bibnamefont
  {He}}, \bibinfo {author} {\bibfnamefont {B.~A.}\ \bibnamefont {Malomed}},
  \bibinfo {author} {\bibfnamefont {F.}~\bibnamefont {Ye}}, \ and\ \bibinfo
  {author} {\bibfnamefont {B.}~\bibnamefont {Hu}},\ }\bibfield  {title}
  {\enquote {\bibinfo {title} {Dynamics of dissipative spatial solitons over a
  sharp potential},}\ }\href {\doibase 10.1364/JOSAB.27.001139} {\bibfield
  {journal} {\bibinfo  {journal} {J. Opt. Soc. Am. B}\ }\textbf {\bibinfo
  {volume} {27}},\ \bibinfo {pages} {1139--1142} (\bibinfo {year}
  {2010})}\BibitemShut {NoStop}%
\bibitem [{\citenamefont {Yin}\ \emph {et~al.}(2011)\citenamefont {Yin},
  \citenamefont {Mihalache},\ and\ \citenamefont {He}}]{Yin_JOSAB_2011}%
  \BibitemOpen
  \bibfield  {author} {\bibinfo {author} {\bibfnamefont {C.}~\bibnamefont
  {Yin}}, \bibinfo {author} {\bibfnamefont {D.}~\bibnamefont {Mihalache}}, \
  and\ \bibinfo {author} {\bibfnamefont {Y.}~\bibnamefont {He}},\ }\bibfield
  {title} {\enquote {\bibinfo {title} {Dynamics of two-dimensional dissipative
  spatial solitons interacting with an umbrella-shaped potential},}\ }\href
  {\doibase 10.1364/JOSAB.28.000342} {\bibfield  {journal} {\bibinfo  {journal}
  {J. Opt. Soc. Am. B}\ }\textbf {\bibinfo {volume} {28}},\ \bibinfo {pages}
  {342--346} (\bibinfo {year} {2011})}\BibitemShut {NoStop}%
\bibitem [{\citenamefont {Liu}\ \emph {et~al.}(2013)\citenamefont {Liu},
  \citenamefont {He},\ and\ \citenamefont {Li}}]{Liu_OE_2013}%
  \BibitemOpen
  \bibfield  {author} {\bibinfo {author} {\bibfnamefont {B.}~\bibnamefont
  {Liu}}, \bibinfo {author} {\bibfnamefont {X.-D.}\ \bibnamefont {He}}, \ and\
  \bibinfo {author} {\bibfnamefont {S.-J.}\ \bibnamefont {Li}},\ }\bibfield
  {title} {\enquote {\bibinfo {title} {Continuous emission of fundamental
  solitons from vortices in dissipative media by a radial-azimuthal
  potential},}\ }\href {\doibase 10.1364/OE.21.005561} {\bibfield  {journal}
  {\bibinfo  {journal} {Opt. Express}\ }\textbf {\bibinfo {volume} {21}},\
  \bibinfo {pages} {5561--5566} (\bibinfo {year} {2013})}\BibitemShut {NoStop}%
\bibitem [{\citenamefont {Cox}\ and\ \citenamefont
  {Matthews}(2002)}]{Cox_2002}%
  \BibitemOpen
  \bibfield  {author} {\bibinfo {author} {\bibfnamefont {S.~M.}\ \bibnamefont
  {Cox}}\ and\ \bibinfo {author} {\bibfnamefont {P.~C.}\ \bibnamefont
  {Matthews}},\ }\bibfield  {title} {\enquote {\bibinfo {title} {{Exponential
  time differencing for stiff systems}},}\ }\href {\doibase
  http://dx.doi.org/10.1006/jcph.2002.6995} {\bibfield  {journal} {\bibinfo
  {journal} {J. Comput. Phys.}\ }\textbf {\bibinfo {volume} {176}},\ \bibinfo
  {pages} {430--455} (\bibinfo {year} {2002})}\BibitemShut {NoStop}%
\bibitem [{\citenamefont {Agrawal}(2001)}]{Agrawal_NWO}%
  \BibitemOpen
  \bibfield  {author} {\bibinfo {author} {\bibfnamefont {G.~P.}\ \bibnamefont
  {Agrawal}},\ }\href@noop {} {\emph {\bibinfo {title} {Nonlinear Fiber
  Optics}}}\ (\bibinfo  {publisher} {Academic Press},\ \bibinfo {address} {San
  Diego},\ \bibinfo {year} {2001})\BibitemShut {NoStop}%
\bibitem [{\citenamefont {Kochetov}\ and\ \citenamefont
  {Tuz}(2018)}]{Chaos_2018}%
  \BibitemOpen
  \bibfield  {author} {\bibinfo {author} {\bibfnamefont {B.~A.}\ \bibnamefont
  {Kochetov}}\ and\ \bibinfo {author} {\bibfnamefont {V.~R.}\ \bibnamefont
  {Tuz}},\ }\bibfield  {title} {\enquote {\bibinfo {title} {Induced waveform
  transitions of dissipative solitons},}\ }\href {\doibase 10.1063/1.5016914}
  {\bibfield  {journal} {\bibinfo  {journal} {Chaos}\ }\textbf {\bibinfo
  {volume} {28}},\ \bibinfo {pages} {013130} (\bibinfo {year}
  {2018})}\BibitemShut {NoStop}%
\bibitem [{\citenamefont {Kochetov}(2019)}]{PD_2019}%
  \BibitemOpen
  \bibfield  {author} {\bibinfo {author} {\bibfnamefont {B.~A.}\ \bibnamefont
  {Kochetov}},\ }\bibfield  {title} {\enquote {\bibinfo {title} {{Mutual
  transitions between stationary and moving dissipative solitons}},}\ }\href
  {\doibase https://doi.org/10.1016/j.physd.2019.01.003} {\bibfield  {journal}
  {\bibinfo  {journal} {Physica D}\ }\textbf {\bibinfo {volume} {393}},\
  \bibinfo {pages} {47--53} (\bibinfo {year} {2019})}\BibitemShut {NoStop}%
\bibitem [{\citenamefont {Caradoc-Davies}(2000)}]{caradoc2000vortex}%
  \BibitemOpen
  \bibfield  {author} {\bibinfo {author} {\bibfnamefont {B.~M.}\ \bibnamefont
  {Caradoc-Davies}},\ }\bibfield  {title} {\enquote {\bibinfo {title} {Vortex
  dynamics in {Bose-Einstein} condensates},}\ }\href@noop {} {\bibfield
  {journal} {\bibinfo  {journal} {Ph. D. thesis}\ } (\bibinfo {year}
  {2000})}\BibitemShut {NoStop}%
\bibitem [{See Supplemental Material at (URL) for animation of waveform
  transformations presented in Figs.~2, 3, 4, and 5 of the
  manuscript()}]{Suppl_Mat}%
  \BibitemOpen
  See Supplemental Material at (URL) for animation of waveform transformations
  presented in Figs.~2, 3, 4, and 5 of the manuscript,\ \href@noop {} {}
  (\bibinfo {year} {2019})\BibitemShut {NoStop}%
\bibitem [{\citenamefont {Wright}\ \emph {et~al.}(2013)\citenamefont {Wright},
  \citenamefont {Blakestad}, \citenamefont {Lobb}, \citenamefont {Phillips},\
  and\ \citenamefont {Campbell}}]{wright2013driving}%
  \BibitemOpen
  \bibfield  {author} {\bibinfo {author} {\bibfnamefont {K.~C.}\ \bibnamefont
  {Wright}}, \bibinfo {author} {\bibfnamefont {R.~B.}\ \bibnamefont
  {Blakestad}}, \bibinfo {author} {\bibfnamefont {C.~J.}\ \bibnamefont {Lobb}},
  \bibinfo {author} {\bibfnamefont {W.~D.}\ \bibnamefont {Phillips}}, \ and\
  \bibinfo {author} {\bibfnamefont {G.~K.}\ \bibnamefont {Campbell}},\
  }\bibfield  {title} {\enquote {\bibinfo {title} {{Driving Phase Slips in a
  Superfluid Atom Circuit with a Rotating Weak Link}},}\ }\href {\doibase
  10.1103/PhysRevLett.110.025302} {\bibfield  {journal} {\bibinfo  {journal}
  {Phys. Rev. Lett.}\ }\textbf {\bibinfo {volume} {110}},\ \bibinfo {pages}
  {025302} (\bibinfo {year} {2013})}\BibitemShut {NoStop}%
\bibitem [{\citenamefont {Eckel}\ \emph {et~al.}(2014)\citenamefont {Eckel},
  \citenamefont {Lee}, \citenamefont {Jendrzejewski}, \citenamefont {Murray},
  \citenamefont {Clark}, \citenamefont {Lobb}, \citenamefont {Phillips},
  \citenamefont {Edwards},\ and\ \citenamefont
  {Campbell}}]{eckel2014hysteresis}%
  \BibitemOpen
  \bibfield  {author} {\bibinfo {author} {\bibfnamefont {S.}~\bibnamefont
  {Eckel}}, \bibinfo {author} {\bibfnamefont {J.~G.}\ \bibnamefont {Lee}},
  \bibinfo {author} {\bibfnamefont {F.}~\bibnamefont {Jendrzejewski}}, \bibinfo
  {author} {\bibfnamefont {N.}~\bibnamefont {Murray}}, \bibinfo {author}
  {\bibfnamefont {C.~W.}\ \bibnamefont {Clark}}, \bibinfo {author}
  {\bibfnamefont {C.~J.}\ \bibnamefont {Lobb}}, \bibinfo {author}
  {\bibfnamefont {W.~D.}\ \bibnamefont {Phillips}}, \bibinfo {author}
  {\bibfnamefont {M.}~\bibnamefont {Edwards}}, \ and\ \bibinfo {author}
  {\bibfnamefont {G.~K.}\ \bibnamefont {Campbell}},\ }\bibfield  {title}
  {\enquote {\bibinfo {title} {Hysteresis in a quantized superfluid
  `atomtronic' circuit},}\ }\href {\doibase 10.1038/nature12958} {\bibfield
  {journal} {\bibinfo  {journal} {Nature}\ }\textbf {\bibinfo {volume} {506}},\
  \bibinfo {pages} {200} (\bibinfo {year} {2014})}\BibitemShut {NoStop}%
\bibitem [{\citenamefont {Yakimenko}\ \emph
  {et~al.}(2015{\natexlab{a}})\citenamefont {Yakimenko}, \citenamefont
  {Bidasyuk}, \citenamefont {Weyrauch}, \citenamefont {Kuriatnikov},\ and\
  \citenamefont {Vilchinskii}}]{yakimenko2015vortices}%
  \BibitemOpen
  \bibfield  {author} {\bibinfo {author} {\bibfnamefont {A.~I.}\ \bibnamefont
  {Yakimenko}}, \bibinfo {author} {\bibfnamefont {Y.~M.}\ \bibnamefont
  {Bidasyuk}}, \bibinfo {author} {\bibfnamefont {M.}~\bibnamefont {Weyrauch}},
  \bibinfo {author} {\bibfnamefont {Y.~I.}\ \bibnamefont {Kuriatnikov}}, \ and\
  \bibinfo {author} {\bibfnamefont {S.~I.}\ \bibnamefont {Vilchinskii}},\
  }\bibfield  {title} {\enquote {\bibinfo {title} {Vortices in a toroidal
  {Bose-Einstein} condensate with a rotating weak link},}\ }\href {\doibase
  10.1103/PhysRevA.91.033607} {\bibfield  {journal} {\bibinfo  {journal} {Phys.
  Rev. A}\ }\textbf {\bibinfo {volume} {91}},\ \bibinfo {pages} {033607}
  (\bibinfo {year} {2015}{\natexlab{a}})}\BibitemShut {NoStop}%
\bibitem [{\citenamefont {Yakimenko}\ \emph
  {et~al.}(2015{\natexlab{b}})\citenamefont {Yakimenko}, \citenamefont
  {Isaieva}, \citenamefont {Vilchinskii},\ and\ \citenamefont
  {Ostrovskaya}}]{yakimenko2015vortex}%
  \BibitemOpen
  \bibfield  {author} {\bibinfo {author} {\bibfnamefont {A.~I.}\ \bibnamefont
  {Yakimenko}}, \bibinfo {author} {\bibfnamefont {K.~O.}\ \bibnamefont
  {Isaieva}}, \bibinfo {author} {\bibfnamefont {S.~I.}\ \bibnamefont
  {Vilchinskii}}, \ and\ \bibinfo {author} {\bibfnamefont {E.~A.}\ \bibnamefont
  {Ostrovskaya}},\ }\bibfield  {title} {\enquote {\bibinfo {title} {Vortex
  excitation in a stirred toroidal {Bose-Einstein} condensate},}\ }\href
  {\doibase 10.1103/PhysRevA.91.023607} {\bibfield  {journal} {\bibinfo
  {journal} {Phys. Rev. A}\ }\textbf {\bibinfo {volume} {91}},\ \bibinfo
  {pages} {023607} (\bibinfo {year} {2015}{\natexlab{b}})}\BibitemShut
  {NoStop}%

\end{thebibliography}

\end{document}